\documentclass[%
 reprint,
 amsmath,amssymb,
aps,
longbibliography,
pra,
]{revtex4-1}

\usepackage{graphicx}
\usepackage{dcolumn}
\usepackage[bookmarks,raiselinks,pageanchor,colorlinks,citecolor=blue,linkcolor=blue,urlcolor=blue,filecolor=blue,menucolor=blue]{hyperref}

\usepackage{mathtools}
\newcommand{\itGamma}{{\it{\Gamma}}} 
\newcommand{\euler}{\mathrm{e}}
\newcommand{\kb}{\mathrm{k}_\mathrm{B}}
\newcommand{\EV}[1]{\left< #1 \right>}
\renewcommand{\vec}[1]{{\boldsymbol{#1}}}
\renewcommand{\d}{\mathrm{d}}

\DeclareSymbolFont{rmlargesymbols}{OMX}{mdbch}{m}{n}
\DeclareMathSymbol{\rmintop}{\mathop}{rmlargesymbols}{82}
\newcommand{\upint}{\rmintop\nolimits}

\newcommand{\diss}{\text{diss}}
\newcommand{\fric}{\text{fric}}
\newcommand{\alkane}{{C\textsubscript{40}H\textsubscript{82}}}
\usepackage{xcolor}

\newcommand{\SICumJI}       {S1} 
\newcommand{\SIgammat}      {S2} 
\newcommand{\SILJ}          {S3} 
\newcommand{\SINaCl}        {S4} 
\newcommand{\SICH}          {S5} 

\begin{document}

\title{Molecular origin of driving-dependent friction in fluids}%

\author{Matthias Post}
\affiliation{Biomolecular Dynamics, Institute of Physics, Albert Ludwigs University, 79104 Freiburg, Germany}%
\author{Steffen Wolf}%
\affiliation{Biomolecular Dynamics, Institute of Physics, Albert Ludwigs University, 79104 Freiburg, Germany}%
\author{Gerhard Stock}%
\affiliation{Biomolecular Dynamics, Institute of Physics, Albert Ludwigs University, 79104 Freiburg, Germany}%
\email{stock@physik.uni-freiburg.de}

\date{\today}%

\begin{abstract}
	The friction coefficient of fluids may become a function of the
	velocity at increased external driving. This non-Newtonian behavior
	is of general theoretical interest as well as of great practical
	importance, e.g., for the design of lubricants. While the effect has
	been observed in large-scale atomistic simulations of bulk liquids,
	its theoretical formulation and microscopic origin is not well
	understood.  Here we use dissipation-corrected targeted molecular
	dynamics, which pulls apart two tagged liquid molecules in the
	presence of surrounding molecules and analyzes this nonequilibrium
	process via a generalized Langevin equation. The approach is based
	on a second-order cumulant expansion of Jarzynski's identity, which
	is shown to be valid for fluids and therefore allows for an exact
	computation of the friction profile as well of the underlying memory
	kernel.  We show that velocity-dependent friction in fluids results
	from an intricate interplay of near-order structural effects and the
	non-Markovian behavior of the friction memory kernel. For complex
	fluids such as the model lubricant \alkane, the memory kernel
	exhibits a stretched-exponential long-time decay, which reflects the
	multitude of timescales of the system.
\end{abstract}

\maketitle

\section{Introduction}

An object driven through a liquid at low to moderate velocities $v$
encounters a drag force $f_{\rm fric} = - v \itGamma$ with the Stokes'
friction factor $\itGamma$. While $\itGamma$ is a constant close to
equilibrium, it may become a function of the velocity at increased
driving. This so-called non-Newtonian behavior of liquids represents
an intriguing topic of fundamental research in rheology and
microrheology \cite{Evans07,Chen10a,Furst21}. Applied to the shearing
of a bulk liquid, the behavior is characterized either as ``shear
thickening'', i.e., an increase of internal friction and thus
viscosity upon shearing, or ``shear thinning'' corresponding to a
decrease of the friction \cite{Cheng11}. The latter is of great
practical importance, e.g., for the design of lubricants which aim to
minimize friction and wear \cite{Bair19}. Non-Newtonian behavior is
believed to result from structural changes within the liquid, such as
the alignment of polymers along the shearing direction
\cite{Huber14,Liu17a,Falk20,Datta21} or from the formation of molecule
clusters \cite{Xu13,Lemarchand15,Maiti16}.

As an alternative to microscopic studies that are based on
large-scale molecular dynamics (MD) simulations, it is instructive to adopt a
coarse-grained description that focuses on the friction experienced by a
driven test molecule. Integrating out all degrees of freedom of the
surrounding molecules, such theoretical formulations yield the
friction in terms of time-dependent molecular correlation functions
\cite{Scheraga55,Vogelsang87,Straub90,Fuchs03,Hummer05,Echeverria14,Straube20}. In
contrast to sliding friction on surfaces whose driving dependence has
been successfully studied using the Prandtl-Tomlinson model
\cite{Mueser11}, the connection between the coarse-grained theoretical
representation and the microscopic origin of velocity-dependent
friction in condensed matter systems is only poorly understood.

\begin{figure}[t!]
	\begin{centering}	
		\includegraphics[width=0.45\textwidth]{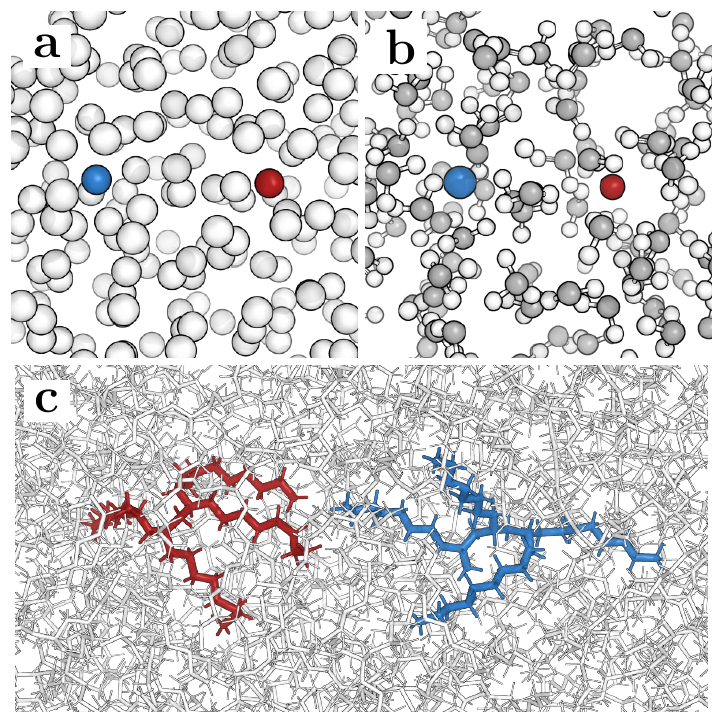}
		\caption{To study the friction in fluids, two tagged molecules (in
			blue and red) are pulled apart using dissipation-corrected targeted
			molecular dynamics (dcTMD) \cite{Wolf18}. Model systems considered
			include (a) a Lennard-Jones liquid, (b) Na$^+$ and Cl$^-$ ions in water,
			and (c) the lubricant \alkane.}
		\label{fig:systems}
	\end{centering}
\end{figure}

In this work we want to amend this understanding by studying
driving-dependent Stokes friction within the theoretical framework of
dissipation-corrected targeted molecular dynamics (dcTMD)
\cite{Wolf18}. The method employs a constraint force $f$ that effects
a moving distance constraint $x = x_0 + v t$ with a constant
velocity $v$ \cite{Schlitter94}, which pulls apart two tagged
molecules (the ``system'') in the presence of the surrounding
molecules (the ``bath'') causing the friction
(Fig.~\ref{fig:systems}). In this way, dcTMD simulations resemble an
active microrheology experiment, where an external force is applied to
a tracer molecule \cite{Furst21}. The approach is based on Jarzynski's
identity\cite{Jarzynski97,Hendrix01,Dellago14}
\begin{align} \label{eq:JI} 
	\Delta G = -  \beta^{-1} \ln \EV{\euler^{-\beta W}},
\end{align}
which estimates the free energy difference $\Delta G$ between two
states of a system from the amount of work $W$ done on this system to
enforce the nonequilibrium process. Here the brackets denote an
ensemble average over statistically independent nonequilibrium
simulations starting from a common equilibrium state, and
$\beta^{-1} = \kb T$ is the inverse temperature. To avoid problems
associated with the poor convergence behavior of the exponential
average (see below), often the second-order cumulant
expansion of Jarzynski's identity is considered,
\begin{align} \label{eq:cumulant} 
	\Delta G = \EV{W} - \EV{W_\diss}
	\approx \EV{W} - \frac{\beta}{2} \EV{\delta W^2} \, ,
\end{align}
where the first term represents the averaged external work performed
on the system, and the second term corresponds to the mean dissipated work
$\EV{W_\diss(x)}$ of the process with
$\delta W = W-\EV{W}$.
Pulling the system from $x_0$ to $x$, we obtain \cite{Wolf18}
\begin{align}
	\EV{W(x)} &= \upint_{x_0}^{x} \d x^\prime \EV{f(x^\prime)}  \, ,  
	\label{eq:Work} \\
	\EV{W_\diss}(x) &= \frac{\beta}{2} \upint_{x_0}^{x} \! \d x^\prime \upint_{x_0}^{x} \! \d x''
	\EV{\delta f(x^\prime) \delta f(x'')} \\
	&= v  \upint_{x_0}^{x} \d x^\prime \itGamma(x^\prime) \, ,  
	\label{eq:Wdiss}
\end{align}
which relates the the dissipated work to the autocorrelation function
of the constraint force $\delta f = f - \EV{f}$, and also provides a means to calculate the
position-dependent friction $\itGamma(x)$ associated with the pulling
process. Using the constraint condition $x = x_0 + v t$ to change the
integration variable from distance $x$ to time $t$, we get
\begin{align}
	\itGamma(x)= \beta \upint_{0}^{t(x)} \! \d{\tau} \EV{\delta f(t) \delta f(\tau)}, \label{eq:GammaNEQ}
\end{align}
which can be directly calculated from the dcTMD simulations. In this
way, $\EV{W_\diss(x)}$ can be considered as dissipation correction of the
work to give the correct free energy, hence the name ``dcTMD''.

Notably, the assumption that the work distribution is well approximated by a Gaussian (in
which case the above cumulant approximation is exact) is in principle
the only condition underlying dcTMD, which might be an advantage
compared to related approaches that calculate friction from pulling
simulations \cite{Park04,Schulz15,Kailasham20}. While this supposition
may be inapplicable to ligand-protein dissociation processes that
provide several exit pathways of the ligand \cite{Wolf20}, we will
show theoretically and numerically that the cumulant approximation is
generally valid for fluid systems.

When we studied the enforced separation of various molecular systems
\cite{Wolf18,Wolf20}, we found that the friction calculated via
Eq.~\eqref{eq:GammaNEQ} exhibits an intrinsic velocity dependence,
whose origin is not apparent from the underlying Markovian Langevin
model. To gain a microscopic understanding of these phenomena, here we
extend the formal basis of dcTMD by deriving a generalized Langevin
equation of the nonequilibrium pulling process
\cite{Zwanzig01,Kawai11,Meyer17,Bingyu18,Lickert21,Schilling21}. This extension
is essential to study the friction in fluid systems, where pulled
molecules and surrounding molecules are of the same size and weight,
therefore impeding a timescale separation required for a Markovian
description. In the Langevin framework, these non-Markovian effects
are described via a two-time memory kernel $K(t,\tau)$ that can be
directly calculated from dcTMD.

Considering various model fluids, from a simple Lennard-Jones liquid
to the complex model lubricant \alkane~(Fig.~\ref{fig:systems}), we
show that the memory kernel may exhibit prominent non-Markovian
behavior such as oscillatory features and a stretched-exponential
decay. As most important effect we find, in a generalization of the
observation that friction may be increased by position constraints
\cite{Daldrop17}, that underlying near-order effects decrease when the
velocity of the pulled molecules approach the timescales of the
bath. That is, both long-time tails of correlation functions
and structural changes of the liquid may cause a velocity-dependence
of the friction in fluids.

\section{Theory}

\subsection{Validation of the cumulant approximation}

As explained above, the main assumption of dcTMD is that the
distribution of the work $W$ resembles a Gaussian to justify the
cumulant approximation \eqref{eq:cumulant}. A common reasoning for
this claim invokes the limit of slow pulling velocities
($v\rightarrow 0$), where the system's response is linear and the work
is given as sum of many independent contributions in
time \cite{Hendrix01,Wolf18}. Extending the argument to independent
contributions in space, here we show that for fluids a Gaussian work
distribution arises as a consequence of the central limit
theorem. Simply speaking, the proposition states that the sum of $N$
independent random variables (with similar distribution) tends towards
a normal distribution for large $N$ \cite{Billingsley95}.

To this end, we consider a fluid with $N + 2$ identical molecules and
pull two molecules from their initial distance $x_0$ to a final
distance $x$. In each pulling simulation, the applied constraint
force $f$ does work comprising the individual contributions $W_i$ of
the other $N$ molecules,
\begin{align} 
	W(\vec{q}_0) = \sum_{i=1}^N W_i \, ,
\end{align}
where $\vec{q}_0$ denotes their initial condition.  (While the work
also depends on time and on the details of the MD force field, these
dependencies are irrelevant for the reasoning here.) 
Since the initial conditions $\vec{q}_0$ are randomly sampled from
equilibrium and the molecules are identical, the other $N$ molecules
have the same probability to scatter with the pulled molecules.
Consequentially, the probability distribution of the work $W_i$ is
identical as well, such that $P(W_i) \sim P(W_j)$ for all molecules
$i,j$. Moreover most contributions $W_i$ are statistically
independent within the ensemble of trajectories, although within a
particular trajectory the work contributions are not necessarily
uncorrelated. These exceptions include molecules in the direct
vicinity of the pulled particles, which may interact in a concerted
manner and therefore result in non-independent work contributions. In
contrast to independent scattering events, they involve at least three
interaction partners, hence these correlations are presumably weak and
may therefore still qualify for the central limit theorem
\cite{Billingsley95}. This should apply to typical interactions
described by a biomolecular force field at ambient conditions, but
excludes fluids close to a critical point \cite{Sornette06}. Hence the
overall work $W$ can be written as sum over $N$ mostly independent
random variables with similar distribution, which is just the
requirement of the central limit theorem.

\subsection{Definition of the friction factor}

In previous work \cite{Wolf18} we derived Eqs.~\eqref{eq:Wdiss} and \eqref{eq:GammaNEQ} that define the dcTMD friction from a Markovian Langevin equation. However, the mere calculation of $\itGamma(x)$ does not require the presumption of a Langevin equation. To show this, we relate frictional forces $f_\fric$ and dissipative work $W_\diss$ via 
\begin{equation}\label{eq:friction}
	f_\fric(x) = - \frac{\partial \EV{W_\diss}}{\partial x} \, ,
\end{equation}
which is valid since the frictional forces are on average the only source of dissipation. While this equation holds for any friction model, we can specifically apply it to Stokes friction, $f_\fric(x) = - v \itGamma(x)$, which directly yields the position-dependent friction
\begin{equation}\label{eq:friction2}
	\itGamma(x) \coloneqq \frac{1}{v} \frac{\partial \EV{W_\diss}}{\partial x} \, .
\end{equation}
Note that this definition of $\itGamma(x)$ is general and does neither rely on a Langevin formalism nor requires a cumulant approximation. Rather by combining Eq.~\eqref{eq:friction2} with Jarzynski's identity, $\Delta G = - \beta^{-1} \ln \EV{ \euler^{-\beta W} } $, we obtain the exact expression
\begin{align}
	\itGamma &= \frac{1}{v} \frac{\partial}{\partial x} \left( \EV{W} - \beta^{-1} \ln \EV{ \euler^{-\beta W} } \right) \label{eq:friction_JI} \\ 
	&= \frac{1}{v} \left( \EV{f} - \frac{ \EV{f \euler^{-\beta W}}}{\EV{\euler^{-\beta W}}} \right)
	\nonumber \\ 
	&= \itGamma_\text{cum} - \frac{\beta^2 v^2}{2} \!\! \upint_0^t \!\!\! \d
	t_2 \! \upint_0^t \!\!\! \d t_1 \! \EV{\delta f(t) \delta
		f(t_2) \delta f(t_1)} + \text{h.o.t.}, \nonumber
\end{align}
where $\itGamma_\text{cum}$ is given in Eq.~\eqref{eq:GammaNEQ}. That is, any deviation from the cumulant approximation results in higher-order correlations of the constraint force contributing to the friction. Unfortunately, these higher-order terms (as well as the full Jarzynski's identity) are notoriously difficult to evaluate due to poor convergence behavior \cite{Hendrix01,Hummer01b}.

\subsection{Generalized Langevin equation}

Given the validity of the cumulant approximation, we have shown above
that the friction can be calculated from the two-time force
autocorrelation function $\EV{ \delta f(t) \delta f(\tau)}$.
Interestingly, the absence of higher-order correlations also allows us
to derive an exact generalized Langevin equation of the nonequilibrium
process. It reads
\begin{equation}\label{eq:GLEneq}
	m \ddot{x}(t) = - \frac{\partial G}{\partial x} - \upint^t_0 \! \d \tau K(t,\tau) \, \dot{x}(\tau)  + \eta(t)  + f(t) \, , 
\end{equation}
and comprises the mean force $- \partial G/\partial x$, a non-Markovian
friction force with memory function $K(t,\tau)$, a stochastic force
$\eta (t)$ due to the associated colored noise with zero mean, and the
external pulling force $f(t)$. 

To derive this equation, we partition the problem into system and
bath, $H=H_S+H_B+H_{SB}$, where the system is
$H_S(x,p) = {p^2}/{2m} + U(x)$, the bath $H_B(\vec{q})$ depends on
phase-space coordinates $\vec{q}$, and the system-bath coupling is
$H_{SB}(x,\vec{q})$. The free energy of the system is given by
\begin{eqnarray}
	G(x) &=&  - \beta^{-1} \ln \upint\! \d \vec{q}\, \euler^{-\beta
		\left[ U(x)+H_{SB}(x,\vec{q}) \right]}
	\nonumber \\
	&\equiv& U(x) + U_{SB}(x).
\end{eqnarray}
Adding the constraint force $f$, we use Hamilton's equation for the
system to find
\begin{equation}\label{eq:appendix_newton}
	m \ddot{x} = -\frac{\partial G}{\partial x} + f_{SB}(x,\vec{q}) + f
\end{equation}
with $f_{SB} = -\partial/\partial x\;(H_{SB}-U_{SB})$.
Due to the constant velocity constraint, $x=x_0+vt$, the total force
vanishes, $m \ddot{x}=0$, and we obtain the work to move the
system from $x_0$ to $x$ via an integration in time
from $0$ to $t$.
Upon averaging over the bath, this yields
\begin{equation}
	\Delta G= \EV{W} + v \upint_{0}^{t} \!\! \d \tau  \, \EV{ f_{SB} } ,
\end{equation}
where the last term is recognized as the dissipative work
$\EV{W_\diss}$. 
Assuming the validity of the cumulant
approximation, moreover, we use Eqs.\ \eqref{eq:Wdiss} and \eqref{eq:GammaNEQ}
to obtain
\begin{align}
	\upint_{0}^{t} \!\! \d \tau \, \EV{ f_{SB} }
	= - \beta v \upint_{0}^{t} \!\! \d t_1 \upint_{0}^{t_1} \!\! \d t_2 \;
	\EV{ \delta f(t_2) \delta f(t_1) }  , 
\end{align}
where we changed the integrated surface from a rectangle to a triangle
to get an additional factor 2. By comparing the integrand on both sides and
defining a fluctuating force via $\eta = f_{SB}- \EV{ f_{SB} }$, we find
\begin{equation} \label{eq:fSB}
	f_{SB}(x,\vec{q}) = - \beta v \upint_{0}^{t} \!\! \d \tau 
	\EV{ \delta f(t) \delta f(\tau)}  + \eta 
\end{equation}
with $\EV{ \eta } = 0$.  Using $v = \dot{x}$,
the insertion of Eq.~\eqref{eq:fSB} in Eq.~\eqref{eq:appendix_newton}
finally yields the desired generalized Langevin
equation~\eqref{eq:GLEneq}.

From the derivation we find that the memory kernel of the generalized
Langevin equation is given by the autocorrelation function of the
constraint force,
\begin{equation}\label{eq:memory_neq}
	K(t_2, t_1) = \beta \EV{ \delta f(t_2) \delta f(t_1) } \, . 
\end{equation}
Hence, dcTMD directly provides the two-time memory kernel accounting
for the non-Markovian behavior of the nonequilibrium pulling
process. This is remarkable, because multi-time memory functions of
nonstationary processes are in general difficult to calculate
\cite{Jung17,Meyer20}. 

Since $\delta f = \eta$, we readily obtain the fluctuation dissipation
theorem
\begin{equation} \label{eq:FDT}
	\beta \EV{ \eta(t_2) \eta(t_1) } = K(t_2,t_1),
\end{equation}
which states that non-white noise $\eta$ causes a finite decay time of the memory kernel $K$. 
Interestingly though, by using Eq.~\eqref{eq:GammaNEQ} 
in combination with $\dot{x}(\tau)=v$, Eq.~\eqref{eq:GLEneq} can be
written in the form of a Markovian Langevin equation
\begin{equation}\label{eq:LEneq}
	m \ddot{x}(t) = - \frac{\partial G}{\partial x} - \itGamma(x)\,
	\dot{x} + \eta(t)  + f(t)  = 0. 
\end{equation}
although the noise is not delta-correlated due to Eq.~\eqref{eq:FDT}.

Owing to the non-stationarity of the underlying process, $K(t_2, t_1)$ generally cannot solely be expressed as a function of the lag time $t_2-t_1$. To nevertheless obtain a simple interpretation of the friction dynamics, we may average the force correlation function over a specific region in $x\in [x_1,x_2]$, where the resulting friction $\itGamma(x)$ varies only little. Using $\tau_i = (x_i-x_0)/v$ ($i=1,2$) to convert this spatial average to a time average, we obtain the one-time memory kernel
\begin{equation}\label{eq:averaged_memory}
	K(\tau) = \frac{\beta}{t_2-t_1}\upint_{t_1}^{t_2} \!\! \d t \, \EV{ \delta f(t+\tau) \delta f(t)} \, .  
\end{equation}
As will be demonstrated in Sec.~\ref{sec:results}, this averaged correlation function adequately illustrates the memory decay of the nonstationary process.

In the equilibrium limit of very slow pulling, we expect that the memory kernel \eqref{eq:averaged_memory} can be compared to the memory function obtained from an unbiased simulation. The latter can be calculated, e.g., by iteratively solving the Volterra equation
\cite{Shin10}
\begin{align} \label{eq:memory_eq}
	m \EV{ \ddot{x}\ddot{x}(t) }  =
	K(t) \EV{ \dot{x}^2 } + \upint_0^t \!\!\d \tau K(\tau) \EV{ \ddot{x}(t-\tau) \dot{x}(t) } ,
\end{align}
which can be derived from the generalized Langevin equation \eqref{eq:GLEneq}.

\subsection{Velocity dependence of the friction}

To focus on the velocity dependence of the friction, we may eliminate the inherent $x$-dependence by averaging $\itGamma(x)$ over some pulling region $[x_1,x_2]$, 
\begin{align}
	\widehat{\itGamma}(v) &\coloneqq \frac{1}{x_2-x_1} \upint_{x_1}^{x_2} \!\! \d x^\prime \itGamma(x^\prime)
	\label{eq:averaged_friction} \\
	&= \frac{1}{v} \frac{ \EV{W_\diss(x_2)} - \EV{W_\diss(x_1)} }{x_2-x_1} ,
\end{align}
where in the second line Eq.~\eqref{eq:friction2} was employed (see also Ref.~\citenum{Kailasham20}). As in the definition of the memory kernel in Eq.~\eqref{eq:averaged_memory}, the averaging window is chosen to represent a characteristic region in $x$.

Since in dcTMD the friction $\itGamma$ is completely determined by the force correlation function $\EV{ \delta f(t_2) \delta f(t_1)}$, this function must also account for the dependence of the friction on the pulling velocity $v$. To study the origin of this effect, we first consider a simple
single-exponential model
\begin{align}\label{eq:expansatz}
	K(t_2,t_1) = \frac{\gamma}{\tau_C} \euler^{-|t_2-t_1|/\tau_C}
\end{align}
with amplitude $\gamma$ and correlation time $\tau_C$. Pulling the system from $x_0$ to a fixed distance $x$ for various velocities $v$, Eq.~\eqref{eq:GammaNEQ} yields the friction
\begin{align}\label{eq:Gv}
	\itGamma =  \gamma \left( 1- \euler^{-v_C/v} \right) \, ,
\end{align}
where $v_C = (x-x_0)/\tau_C$.
The velocity dependence is characterized by the ratio $v_C/v$, which relates the response rate of the bath to the velocity of the constrained particles. For low pulling velocities with $v_C/v \gg 1$, we find that $\itGamma = \gamma$. In the opposite case of high pulling velocities, the upper limit $t=(x-x_0)/v$ of the integral in Eq.~\eqref{eq:GammaNEQ} becomes shorter than the decay time $\tau_C$ of the autocorrelation function, such that the integral does not reach its maximum value. This reflects the fact that for a given pulling range $x - x_0$ the system is pulled too fast to access all dissipation channels present in the bath. The resulting velocity dependence is generic, in the sense that the effect occurs independent of the considered system as long as the pulling velocity is high enough.

When we average Eq.~\eqref{eq:Gv} over some pulling region $[x_0, x]$ as in Eq.~\eqref{eq:averaged_friction}, the averaged friction $\widehat{\itGamma}(v)$ also shows this generic behavior. Averaging over a region where $\itGamma$ already converged to $\gamma$, we find $\widehat{\itGamma}=\gamma$. In general, on the other hand, we obtain
\begin{align}
	\widehat{\itGamma}(v) = \gamma \left( 1- \frac{v}{v_C}\left( 1 - 
	\euler^{-v_C/v}\right)\right) \label{eq:model_one_exp} ,
\end{align}
which again depends on the pulling velocity.

We have also considered more complex forms of the autocorrelation
function instead of Eq.~\eqref{eq:expansatz}, such as bi-exponential
and stretched-exponential models reflecting the presence of multiple
timescales of the bath \cite{Hamm06}. While they too lead to explicit
expressions for $\itGamma$ and $\widehat{\itGamma}$ (see SI Methods),
in essence they describe the same generic effect as 
discussed above. This indicates that more involved cases, such as a
velocity dependence due to near-order effects, cannot be described by
exponential models with constant amplitudes and decay times.

\section{Methods}

All simulations were performed using Gromacs version 2016.3
\cite{Abraham15}, employing the implemented PULL code for
dcTMD. Analysis of the dcTMD trajectories was carried out using our
dcTMD Python package that is available at
\url{www.moldyn.uni-freiburg.de}.

\subsection{Lennard-Jones liquid.}

The Lennard-Jones model consists of 800 argon atoms in a rectangular box
of 3.4\,nm width and periodic boundaries. Starting with a
configuration in which two argon atoms exhibit a distance of
$x = 0.35$~nm, we perform 10 ns of simulation with 1 fs time-step
under $NVT$ equilibrium conditions. We use the Bussi thermostat
\cite{Bussi07} at 80~K with a coupling time constant $\tau \!=\! 0.1$~ps,
a van der Waals cut-off of 0.85~nm, and a distance constraint on $x$
to obtain 10\,000 equidistant system snapshots. These snapshots serve
as starting configurations of 0.1~ns long $NVT$ runs, which keep two
tagged atoms at a fixed distance $x$, and employ redistributed
velocities to generate an initial Boltzmann distribution for the
subsequent nonequilibrium simulations. Pulling the two tagged atoms
from $x\!=\!0.35$ to 1.65~nm, we performed 5\,000 dcTMD runs for
velocities up to $0.005$~nm/ps, and 10\,000 runs for higher velocities.
The dcTMD simulations were compared to a 1~ns long unbiased
$NVT$ equilibrium simulation with 10~fs resolution.

\subsection{NaCl in water.}

Extending the studies of Ref.~\citenum{Wolf18}, we performed $NPT$
simulations using 893 water molecules plus a Na$^+$ and a Cl$^-$ ion
in a cubic box of 3~nm side length. Employing a 1\,ns simulation of
equilibration at fixed ion distance $x = 0.265$~nm, we generated
1000 starting points. After performing a 0.1~ns equilibration run for
each point, we picked 10 frames of each run, each of which was again
equilibrated for 10~ps assuming random velocities, to end up with in
total 10\,000 starting configurations. To obtain data for long ion
distances, we pulled up to $x = 1.265$~nm using velocities
$v = 0.0001$, $0.001$, $0.01$ and $0.1$~nm/ps, and a time step of
1~fs. To better resolve velocity-dependent friction features, we
employed additional pulling velocities, but pulled only until
$x = 0.865$~nm. In this way, we simulated 1000 trajectories for
velocities $v \le 0.0005$~nm/ps, 5\,000 for $0.001$~nm/ps $\le v \le 0.006$~nm/ps, and 10\,000 for higher velocities.
We compared the dcTMD results to 50 unbiased simulations of 1~ns, writing
out structural snapshots every 10~fs.

\subsection{Lubricant \alkane.}

All simulations are based on structures from Ref.~\citenum{Falk20}
with 175 \alkane~molecules in a cuboid of size
$5\!\times 6.5\!\times\!8$~nm$^3$ and periodic boundary
conditions. Force field parameters were generated using antechamber
\cite{Wang06b} and acpype \cite{SousadaSilva12}, employing OPLS-AA atom
types \cite{Dodda17} and AM1/BCC charges
\cite{Jakalian00,Jakalian02}. The structures were minimized using
steepest descend. Equilibration was performed in the $NVT$ ensemble
for 5~ns (1~fs time step), using the Bussi thermostat at 600~K
($\tau = 0.2$~ps). For Coulomb interactions, we used the Particle
Mesh Ewald summation method \cite{Darden93} with a real-space cut-off
of 1.3~nm, and a van der Waals cut-off of 1.3~nm.
Production runs consisted of one $NVT$ run of 100~ns to generate
initial configurations for the subsequent pulling
simulations. Starting from 1000 initial configurations, two
hydrocarbons at a center-of-mass distance $x_0 \approx 0.6$~nm were
picked as tagged particles. After pulling to the exact position if
necessary, the systems were again equilibrated with fixed $x$ for
100~ps. Using a pulling range from $x = 0.6$ to 2.1~nm, the
resulting configurations were used as starting conditions for the
following dcTMD simulations:
200 runs for $v \le 0.00075$~nm/ps, 500 runs for 0.0015~nm/ps $\le v
\le 0.005$~nm/ps, and 1000 runs for higher velocities.
Moreover, we performed an 1~ns-long unbiased simulation using a 1~fs
resolution.

\subsection{Computation of friction and memory kernels.}

As described above, we perform for each system and pulling velocity a
number of nonequilibrium dcTMD simulations, from each of which we record
the constraint forces $f(t)$ at a temporal resolution of 1~fs. By
integrating $f$ via the trapezoidal rule, we calculate the work $W$
and its mean $\EV{W}$ and variance
$\EV{\delta W^2}$ for all considered distances $x$, which
yields the free energy profile $\Delta G(x)$ and the dissipated work
$W_\diss(x)$ via Eq.~\eqref{eq:cumulant}.

By numerical differentiation of $\EV{W_\diss(x)}$, we then calculate
the friction profile $\itGamma(x)$ via Eq.~\eqref{eq:friction2}. Since
the derivative is prone to large fluctuations, we employ an averaging
window $\Delta x$, using $\Delta x = 0.026$~nm for the Lennard-Jones model, 
$0.02$~nm for NaCl and 0.15~nm for \alkane. In this way, fluctuations are
minimized, while main features of the friction profile are still
resolved. While the free energy $\Delta G$ is given as the difference
of two large numbers in dcTMD [Eq.~\eqref{eq:cumulant}], the friction
$\itGamma$ can be directly obtained from $\EV{W_\diss(x)}$ and is
therefore less prone to statistical errors than $\Delta G$ in the
trajectory average.

As additional test of the cumulant approximation, we also calculated
$\widehat{\itGamma}$ directly via Jarzynski's identity
[Eq.~\eqref{eq:friction_JI}]. While cumulant and Jarzynski results
agree well in general (Fig.~\SICumJI), for large velocities we still
find well-known convergence issues of Jarzynski's identity
\cite{Hendrix01,Hummer01b}.

The dcTMD memory kernel $K(t)$ is directly computed from the
constraint force trajectories $\delta f (t)$ according to
Eq.~\eqref{eq:averaged_memory}.
To study the generic onset of the friction discussed in Eq.~\eqref{eq:Gv}, 
it is instructive to integrate the memory kernel,
$\gamma(t) = \upint^{t}_0 \! \d \tau \, K(\tau)$, and consider the
convergence of $\gamma(t)$ for various pulling velocities (Fig.\
\SIgammat).

To compare the dcTMD results to the equilibrium memory kernel obtained
from unbiased simulations, we iteratively solved
Eq.~\eqref{eq:memory_eq} for $K(t)$ as described in Ref.~\citenum{Shin10}. 
As this requires the accurate calculation of velocities
and forces by finite differences, we used a short time step of 10~fs
for the Lennard-Jones model and NaCl and 1~fs for
\alkane. Using a 10~fs time step, we had to correct for the
artificial force spike at $t = 0$ by employing a parabolic fit to
the subsequent time frames. Moreover, since the estimator may fail at
the long-time tail of the memory kernel, we performed a fit of the
tail to correct for the noise and to extrapolate at longer times. That
is, for the Lennard-Jones model we fit a mono-exponential function to
the results obtained for $0.8-1.4$~ps, in order to extend the kernel
from 1.4~ps onward. For NaCl we performed an exponential fit from 0.4
to 2.0~ps to connect at 0.5 ps, and for \alkane~we used a
stretched-exponential fit from 2.2 to 20~ps to continue from 10~ps.

\section{Results and discussion}
\label{sec:results}

\subsection{Friction and memory of a Lennard-Jones liquid.}

\begin{figure}[h!]
	\begin{centering}	
		\includegraphics[width=0.45\textwidth]{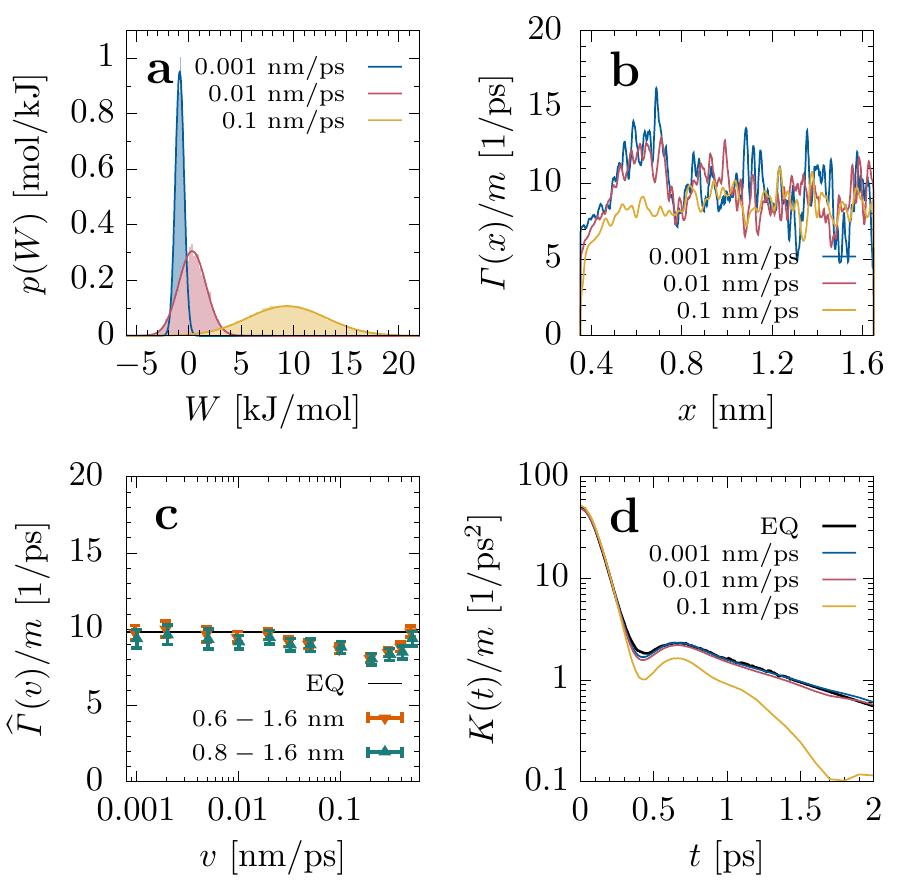}
		\caption{Characterization of the friction of a Lennard-Jones liquid as
			obtained from dcTMD simulations using various pulling velocities
			$v$. (a) Distribution of the work $W$ performed on the system, (b)
			position-dependent friction $\itGamma(x)$
			[Eq.~\eqref{eq:friction2}], (c) position-averaged friction
			$\widehat{\itGamma}(v)$ [Eq.~\eqref{eq:averaged_friction}] evaluated
			for two averaging windows with error bars corresponding to a
			bootstrap estimate of the 95\% confidence interval, and (d) friction
			memory kernel [Eq.~\eqref{eq:averaged_memory}]. EQ indicates
			equilibrium results obtained from unbiased MD simulations.}
		\label{fig:argon}
	\end{centering}
\end{figure}

To demonstrate how dcTMD works in practice, we start with a
Lennard-Jones liquid as a system of minimal microscopic
complexity. The idea is to pull apart two tagged Lennard-Jones
particles, which results in a partitioning into a ``system'' of two
spheres that are embedded in a ``bath'' consisting of the remaining
particles (Fig.~\ref{fig:systems}a). Since the constrained particles
are of the same size and weight as the bath particles, the model
challenges the ansatz of a Markovian Langevin equation employed in the
original formulation of dcTMD \cite{Wolf18}, but should be well
represented by the generalized Langevin equation [Eq.~\eqref{eq:GLEneq}].
As the main assumption of dcTMD is a Gaussian work distribution $P(W)$
to justify Eq.~\eqref{eq:cumulant}, we first show in Fig.~\ref{fig:argon}a 
that this is indeed the case for all considered
pulling velocities $v$. Comparing $P(W)$ directly to the corresponding
normal distribution (Fig.~\SILJ a), we only find deviations at the
poorly sampled tails ($|W| \gtrsim 3 \sigma$) of the
distribution. Moreover, we checked that dcTMD reproduces the
well-known radial distribution function and corresponding free energy
profile of a Lennard-Jones liquid for velocities up to 0.01~nm/ps (Fig.~\SILJ b).

We are now in a position to discuss the friction properties of the
Lennard-Jones liquid, see Fig.~\ref{fig:argon}. To represent the
magnitude of the friction in terms of a damping rate with unit 1/ps,
in all figures we divided the friction by the reduced mass $m$ of the
considered system.  Depicted as a function of the position, the
friction $\itGamma(x)$ defined in Eq.~\eqref{eq:friction2} starts at
$\itGamma(x_0 = 0.35\,{\rm nm})=0$. The initial steep rise of
$\itGamma(x)$ reflects the generic onset of the friction discussed in
Eq.~\eqref{eq:Gv}. While this effect becomes more important with
increasing velocity, $\itGamma(x)$ already plateaus for
$x \approx 0.5$~nm even for 0.1~nm/ps, meaning that this type of velocity
dependence is not relevant here. For long inter-particle distances,
$x \gtrsim 0.8$~nm, the friction is found to fluctuate around a
roughly constant value.

As an unexpected feature, we note a significant maximum of
$\itGamma(x)$ at $x\approx 0.6$~nm for small velocities. This is a
consequence of the fact that this distance is just large enough for a
third particle to shortly interact with both pulled particles. As can
be seen from the particle distribution around the pulled particles
(Fig.~\SILJ c), the structural arrangement is similar to a small cluster of
crystallized Lennard-Jones spheres. Caused by pulling-induced
suppression of fluctuations of the constrained particles, the effect
does not occur in the bulk liquid \cite{Daldrop17}. Thus, if $x$ is
changed slowly enough, interactions with the bath may take place that
would not occur in the unconstrained case.

To further study the velocity dependence of the friction, we next
consider the position-averaged friction $\widehat{\itGamma}(v)$
[Eq.~\eqref{eq:averaged_friction}] shown in Fig.~\ref{fig:argon}c.
Performing averages with and without including the maximum of
$\itGamma(x)$ around $x = 0.6$~nm is found to hardly change
$\widehat{\itGamma}(v)$. Rather we see that overall the
$\widehat{\itGamma}(v)$ is well approximated by its equilibrium value
calculated from Eq.~\eqref{eq:memory_eq}.
Starting with $v = 0.01$~nm/ps, however, we observe a small gradual
decrease of the friction up to $v = 0.2$~nm/ps, where the probe
particles approach the mean thermal velocity
$\EV{v} = \sqrt{8 \kb T / (\pi m)} \approx
0.29$~nm/ps. That is, the pulled particles experience less drag if they
move with similar speed as the surrounding particles, thus
minimizing the collisions with the environment. The
effect vanishes for even larger pulling velocities, when the
constrained particles move faster than the average bath velocity.

As system and bath particles are identical and therefore do
not exhibit a timescale separation, it is interesting to study the
memory kernel $K(t)$ that reports on the resulting non-Markovian
behavior. We first discuss the equilibrium friction kernel, which was
calculated from an unbiased simulation using
Eq.~\eqref{eq:memory_eq}. As shown in Fig.~\ref{fig:argon}d, $K(t)$
undergoes a rapid initial decay leading to a recurrence at
$t = 0.6$~ps, before it decreases exponentially on a timescale of
$\approx 1$~ps. Here the initial decay and the recurrence account for
the interaction of two colliding particles with an average interaction
time of $\approx 0.6$~ps, while the long-time decay reflects the average
time between two scattering events.  We note that a phenomenological
bi-exponential ansatz of the memory kernel (in the spirit of
Eq.~\eqref{eq:expansatz}) would qualitatively reproduce these results,
with the exception of the recurrence which reflects the microscopic
collision process.

Comparing the equilibrium friction kernel to the nonequilibrium
results obtained from dcTMD [Eq.~\eqref{eq:averaged_memory}],
Fig.~\ref{fig:argon}d reveals almost perfect agreement of equilibrium
and dcTMD results for velocities up to 0.01~nm/ps.  For
$v=0.1$~nm/ps, on the other hand, the dcTMD memory kernel decays
significantly faster compared to the equilibrium kernel, which is in
line with the decrease of $\widehat{\itGamma}(v)$ in
Fig.~\ref{fig:argon}c. Evidently, the latter behavior cannot be
reproduced with a bi-exponential memory kernel with constant decay
times.

\subsection{Dissociation of NaCl in water}

\begin{figure}[h!]
	\begin{centering}	
		\includegraphics[width=0.45\textwidth]{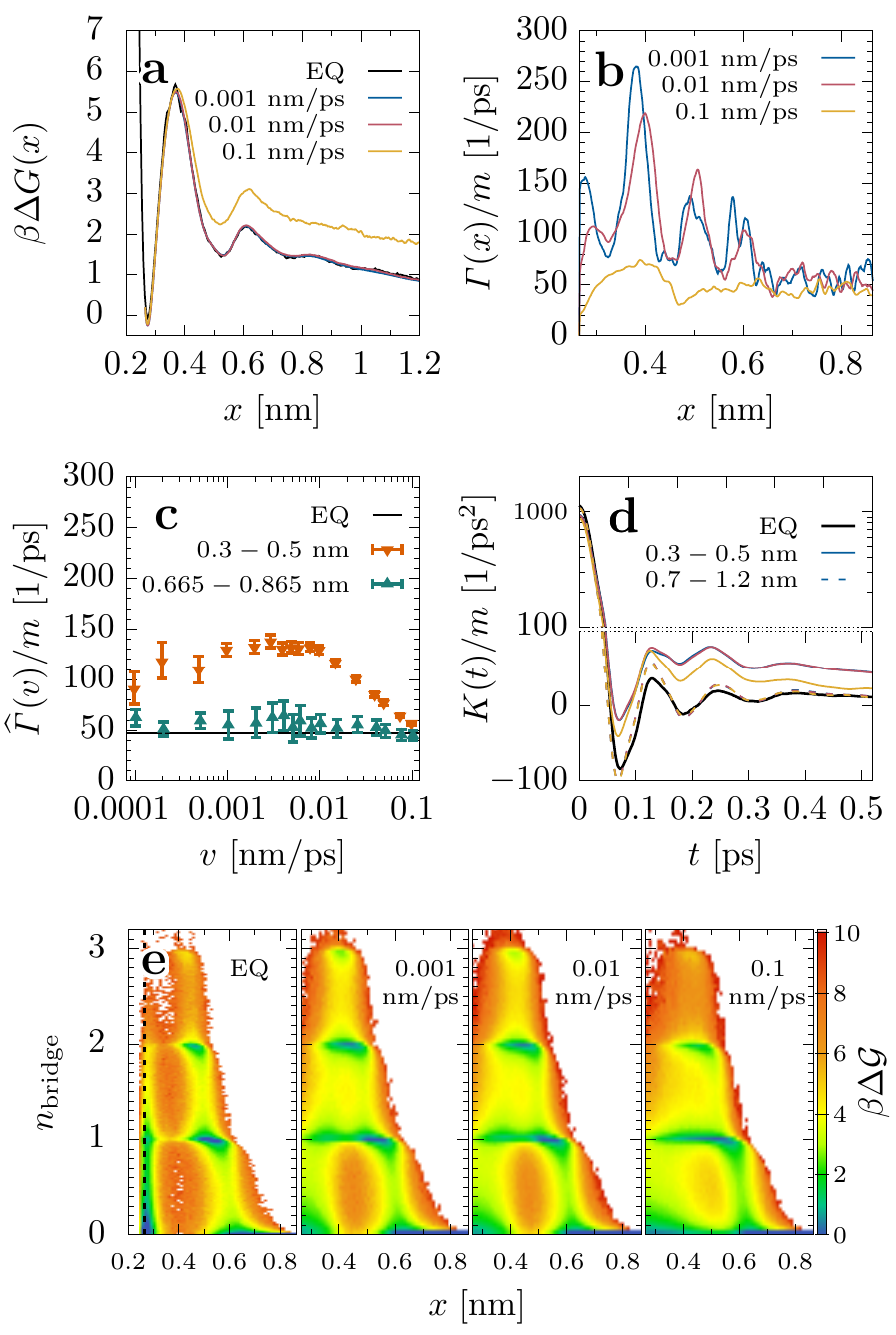}
		\caption{Complex friction behavior of NaCl in water. Shown are (a)
			free energy $\Delta G(x)$ and (b) friction profiles $\itGamma(x)$ as
			a function of the interionic distance $x$, (c) position-averaged
			friction $\widehat{\itGamma}(v)$ evaluated for two averaging windows
			with error bars corresponding to a bootstrap estimate of the 95\%
			confidence interval, and (d) associated friction memory kernels
			which are depicted on a logarithmic scale for $K/m \ge 100/$ps$^2$
			and on a linear scale otherwise. Color code is as in (b). (e)
			Probability distribution that the ions at distance $x$ are connected
			via $n$ water molecules.}
		\label{fig:nacl}
	\end{centering}
\end{figure}

The enforced ion dissociation of NaCl in water can be viewed as as
idealized microrheological experiment that probes the interactions
experiences by the ions \cite{Furst21}.  While we still consider only
two particles, the situation turns out to be much more complex than a
Lennard-Jones model, which is because of the electrostatic interaction
between the Na$^+$ and Cl$^-$ ions and the structured and dynamical
water environment. This is already evident from the free energy
profile $\Delta G(x)$ along the interionic distance $x$
(Fig.~\ref{fig:nacl}a), whose first maximum at $x \approx 0.4$~nm
corresponds to the binding-unbinding transition of the two ions, while
the second smaller maximum at $x \approx 0.6$~nm reflects the
transition from a common to two separate hydration shells
\cite{Mullen14}. As the enforced separation of the ions again results
in a Gaussian work distribution for all considered pulling velocities
(Fig.~\SINaCl a,b), the supposedly complex friction behavior of
solvated NaCl is nevertheless amenable to dcTMD modeling.

Figure~\ref{fig:nacl}b shows the resulting friction profile
$\itGamma(x)$ for various pulling velocities, which is indeed quite
structured in the region of the two solvation shells
($x \lesssim 0.7$~nm), where the maxima at 0.39 and 0.59~nm
correlate well with the maxima found for $\Delta G(x)$. Hence these
features are associated with the formation of hydration shells shared between
the ions \cite{Wolf18}, which can also be seen in two-dimensional
density plots of the system (Fig.~\SINaCl c). As for the Lennard-Jones model
(Fig.~\ref{fig:argon}b), the structural features vanish for larger
pulling velocities.  Moreover, the generic onset of the friction
[Eq.~\eqref{eq:Gv}] again does not play a mayor role, as
$\itGamma(x)$ reaches a plateau for $\approx 0.4$~nm even at the highest
considered velocity ($v = 0.1$~nm/ps). For distances larger than 0.7~nm,
$\itGamma (x)$ converges to a common value for all velocities.  Here,
the friction is dominated by collisions of the ions with water
molecules from their own hydration shells \cite{Wolf18}.

Because of the large difference of $\itGamma (x)$ for small (0.3~nm
$\leq x \leq$ 0.5~nm) and large (0.665~nm $\leq x \leq$
0.865~nm) distances, it is instructive to separately consider these
two averaging windows when we study the velocity dependence of the
averaged friction $\widehat{\itGamma}(v)$ (Fig.~\ref{fig:nacl}c). As
expected from Fig.~\ref{fig:nacl}b, we find that for large distances
$\widehat{\itGamma}(v)$ is constant, and in fact coincides well with
its equilibrium value calculated from Eq.~\eqref{eq:memory_eq}. For
short distances, on the other hand, we find a significant velocity
dependence, i.e., the friction is enhanced by up to a factor 3 for
moderate velocities ($v \lesssim 0.01$~nm/ps) and decreases to its
equilibrium value for higher velocities. Similar to the Lennard-Jones
liquid, the enhancement is caused by constraint-induced fluctuations
that do not occur at equilibrium. (Indeed, unconstrained
calculations of $\itGamma (x)$ revealed a structureless friction
profile \cite{Lickert20}.) Also similarly, the friction decreases
when the pulling approaches the picosecond timescale of the reordering
of the solvation shells, causing the ions to experience less
drag. In contrast to the Lennard-Jones model, however, these effects
are much larger (factor 3 vs.\ 1.2) for the water-solvated ion system.

The above finding of a velocity-dependent reordering of common ion
hydration shells can be nicely illustrated by considering the
probability distribution that the ions at distance $x$ are connected
via $n = 0,1,2,3$ water molecules (Fig.~\ref{fig:nacl}e)
\cite{Mullen14,Wang22}. In the unbiased case, we find typically zero
to one bridging water molecules at the minimum ($x\approx 0.28$~nm), while
up to three bridging waters are found during dissociation or
association ($x\approx 0.35$~nm). For the constrained distribution
\cite{Post19}, we indeed observe a bias towards a higher number of
connecting water molecules than present in the unbiased case,
indicating altered dynamics within the hydration shells. For fast
pulling velocities, the number of bridging water molecules again
gradually becomes smaller -- up to a point, where an absence of
bridging water molecules during the transition becomes viable as well.

As the Na$^+$ and Cl$^-$ ions and the surrounding water molecules are
of similar mass and thus move on a similar timescale, we again expect
non-Markovian behavior of the system. Figure~\ref{fig:nacl}d shows the
corresponding friction memory kernel obtained form unbiased
equilibrium simulations and dcTMD simulations. In the equilibrium
case, the memory kernel exhibits an rapid ($\approx 25$~fs) initial
decay, which is followed by oscillatory features with a period of
$\approx 120$~fs, that are damped on a 1~ps timescale. While the
initial decay reflects the timescale of direct ion-water collisions
\cite{Wolf18}, the long-time decay of the kernel agrees well with the
average lifetime (0.77~ps) of a water molecule in an ion hydration
shell (Fig.~\SINaCl d). As discussed in Ref.~\citenum{Meyer20}, the
oscillatory features correspond to damped ion-water oscillations.

When using dcTMD to calculate the memory kernel for finite pulling
velocities, we again consider averaging windows for short and long
distances [Eq.~\eqref{eq:averaged_memory}]. In the latter case, the
dcTMD kernel closely recovers the unbiased results for all
velocities. Considering short distances, on the other hand, the
amplitude of the long-time decay is significantly increased, which
again is a consequence of the above discussed reordering of the
solvation shells. A similar effect was observed for the interaction of
constrained van der Waals particles within bulk water
\cite{Daldrop17}.

\subsection{Friction and long-time memory of \alkane} 

\begin{figure}[h!]
	\begin{centering}	
		\includegraphics[width=0.45\textwidth]{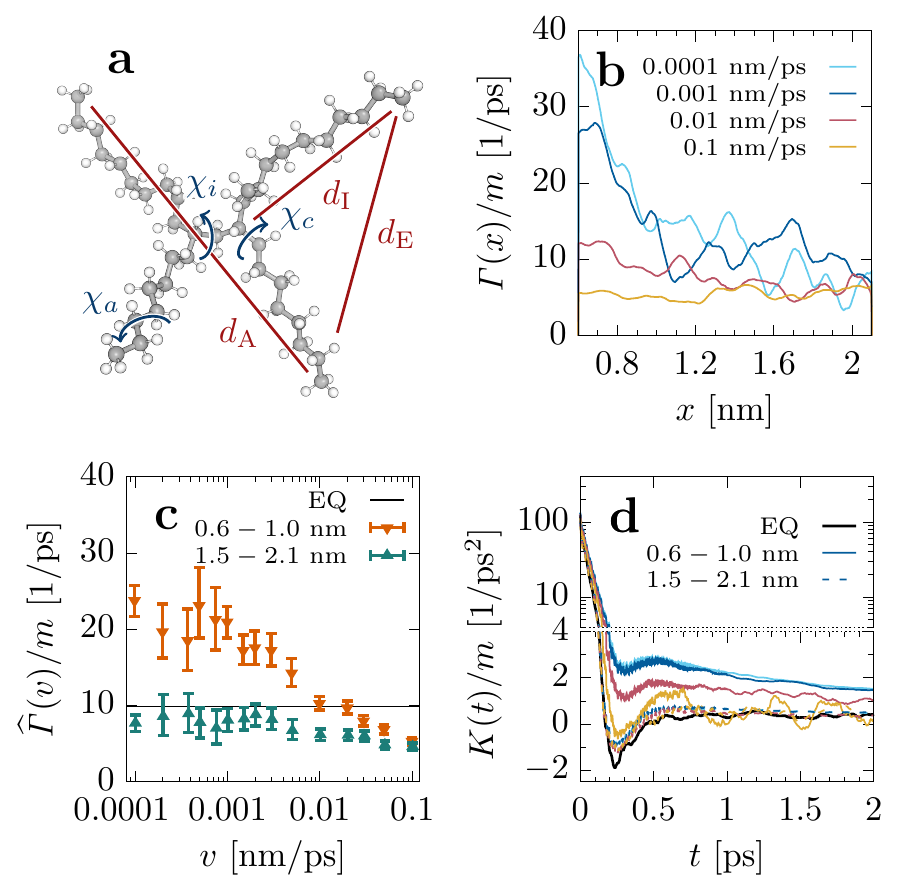}
		\caption{Friction of a hydrocarbon lubricant. (a) Structure of
			\alkane, indicating various inter-atom distances $d_j$ and
			backbone dihedral angles $\chi_j$. (b) Friction profiles
			$\itGamma(x)$ as a function of the center-of-mass distance $x$,
			obtained for various pulling velocities $v$. (c) Position-averaged
			friction $\widehat{\itGamma}(v)$ evaluated for two averaging
			windows. Error bars correspond to a bootstrap estimate of the 95\%
			confidence interval. (d) Memory kernel $K(t)$, shown on a
			logarithmic scale for $K/m \ge 4$~ps$^{-2}$ and on a linear scale
			otherwise. Color code is as in (b).}
		\label{fig:c40}
	\end{centering}
\end{figure}

The hydrocarbon \alkane~represents a model lubricant with a
well-established shear-velocity dependence \cite{Falk20}. To probe the
friction and its dependence on velocity, we pull two tagged
\alkane~molecules apart along their centers of mass. Due to
their four-armed starfish-like structure (Fig.~\ref{fig:c40}a) and
relatively dense packing (Fig.~\ref{fig:systems}c), the tagged
and surrounding alkane molecules interact in a vastly complex,
intertwined fashion. While we may expect an intricate behavior of the
friction and the memory kernel, pulling simulations again yield a
Gaussian work distribution (Fig.~\SICH a,b) which facilitates a dcTMD
analysis.

The dcTMD friction profile $\itGamma(x)$ shown in Fig.~\ref{fig:c40}b
exhibits a clear position and velocity dependence. While the data are
more noisy than for the simpler systems discussed above, we again find
for slow velocities that the friction is higher at short distances,
while at large distances the slow-velocity friction approaches the
high-velocity results. Accordingly, when averaged over short
distances, the friction $\widehat{\itGamma}(v)$ decreases with
increasing velocity (Fig.~\ref{fig:c40}c). When averaged over long
distances, on the other hand, $\widehat{\itGamma}(v)$ remains roughly
constant and close to its equilibrium value up to $\approx 0.003$~nm/ps, from
where is starts to drop slightly. Since the arms of the two pulled
alkane molecules can be in contact even for a center-of-mass distance
of 2.0~nm, this velocity dependence at long distances may either be
attributed to remaining direct interactions or to near-ordering
effects as found for the other two systems. In fact, nonequilibrium MD
simulations of similar systems revealed an overall alignment of the
lubricants along the shearing direction, which also results in a
decreased radius of gyration\cite{Liu17a,Falk20}. Unlike
to the simpler systems, \alkane~moreover exhibits a generic
velocity dependence [Eq.~\eqref{eq:Gv}], because the underlying
friction memory kernel decays too slow to be completely integrated by
fast pulling velocities (Fig.~\SIgammat).

To discuss the rather complex behavior of this memory kernel $K(t)$
shown in Fig.~\ref{fig:c40}d, we first focus on the equilibrium
results for $K(t)$. Initially the kernel exhibits a decay until
$\approx 0.2$~ps, which corresponds to thermal motions and collisions of
the two constrained molecules. Following a recurrence at
$t\approx 0.5$~ps, the kernel shows a slow decay, which is best fitted
to a stretched-exponential function ($\sim e^{(t/\tau)^\alpha}$
with $\alpha\approx 0.2$).
To obtain dcTMD results, we again averaged the memory kernel over
either short or long distances [Eq.~\eqref{eq:averaged_memory}]. 
While in the latter case the dcTMD kernel
closely recovers the equilibrium results for all velocities, short
distances result in an significant increase of the amplitude of the
long-time decay for slow velocities, which is the main origin of the
velocity dependence of $\widehat{\itGamma}(v)$ in Fig.~\ref{fig:c40}c.

While a mono-exponential long-time decay of the memory kernel found
for the simpler systems suggests a single main timescale of the bath,
the stretched exponential form of $K(t)$ in Fig.~\ref{fig:c40}d is
indicative of bath dynamics on a multitude of timescales
\cite{Hamm06}. To elucidate the underlying microscopic mechanism, it
is instructive to illustrate the motion of \alkane~along
representative internal coordinates as indicated in
Fig.~\ref{fig:c40}a. When we calculate the equilibrium
autocorrelation function of the these coordinates, we find that
motions involving a single arm of \alkane, such as the
dihedral angle $\chi_a$ or the distance $d_I$, decay on a picosecond
timescale (Fig.~\SICH e,f). Distances connecting the ends of two arms
such as $d_A$ and $d_E$, are found to decay on tens of picosecond,
while the internal rotation of one or two arms described by dihedral
angles $\chi_C$ and $\chi_i$ takes about 100~ps (Fig.~\SICH e,f). 
Hence it is the combination of the above described bath motions occurring on
timescales across three orders of magnitude that yield the stretched
exponential long-time decay of the memory kernel.

\section{Conclusions}

To gain a microscopic understanding of the phenomena leading to
velocity-dependent friction in fluids, we have employed
dissipation-corrected targeted molecular dynamics (dcTMD), which pulls
apart two tagged molecules (the ``system'') in the presence of the
surrounding molecules (the ``bath''). Since system and bath particles
are identical in fluids and therefore do not exhibit a timescale
separation, we have extended the theoretical framework of dcTMD to
account for the resulting non-Markovian behavior of the system.
dcTMD was shown to directly provide the nonequilibrium memory kernel
$K(t_2,t_1)$ [Eq.~\eqref{eq:memory_neq}], from which the mean dissipated
work $\EV{W_\diss(x)}$ and the associated friction profile
$\itGamma(x)$ along the pulling path $x$ [Eq.~\eqref{eq:friction2}]
is readily calculated. Moreover, we have shown that the nonstationary
kernel $K(t_2,t_1)$ can be related to an equilibrium memory kernel
$K(t)$ [Eq.~\eqref{eq:averaged_memory}], and compared this result to
a kernel obtained from a Volterra equation derived from unbiased
simulations.
Because dcTMD rests on the assumption 
that the distribution of the work $W$ is well approximated by a Gaussian
(to justify the cumulant approximation in Eq.~\eqref{eq:cumulant}),
we have validated this condition for all considered fluids and pulling
velocities. As a consequence, the nonequilibrium process is 
completely described by a two-time autocorrelation function and thus 
exactly modeled by a generalized Langevin equation~\eqref{eq:GLEneq}. 
As the cumulant approximation represents the
only assumption underlying dcTMD, this result is remarkable because it
suggests that dcTMD should be a valuable approach to characterize the
friction behavior of even complex fluids.

Considering three fluids, from a simple Lennard-Jones liquid, via
solvated NaCl to the complex lubricant \alkane, we have
identified near-order structural effects as main reason for the
apparent velocity dependence of the friction for all systems. It
starts with the observation that the friction experienced by
position-constrained molecules generally increases compared to
unconstrained molecules \cite{Daldrop17}. This effect typically occurs
at short distances of the two constrained molecules, where the
molecules either interact directly or via the surrounding
molecules. At long distances, these interactions vanish and the
friction decreases to its equilibrium value.  Being pulled apart with
slow velocity, the molecules therefore experience enhanced friction as
long as these near-order effects prevail. In a similar vein,
perturbation of local structures around a pulled probe has been linked
to the onset of shear thinning in a theoretical study on microrheology
\cite{Gazuz09}.  In the case of high pulling velocity similar to the
speed of the surrounding particles, the particles experience less
drag, because they minimize the collisions with the environment.

Although the above described mechanism is similar for all considered
systems, the resulting effect on the friction differs
significantly, reflecting the quite different near-order effects of
the fluids. While the friction in a short-range Lennard-Jones liquid
hardly depends on the pulling velocity at all (Fig.~\ref{fig:argon}),
the well-defined hydration shells of Na$^+$ and Cl$^-$ ions in water
give rise to a very structured friction profile $\itGamma(x)$ and a
short-distant enhancement of the friction by a factor 3
(Fig.~\ref{fig:nacl}).
The four-armed lubricant \alkane, on
the other hand, may exhibit direct interactions of the two pulled
alkanes up to 2~nm, which enhances the friction by a factor 2 at low
velocities (Fig.~\ref{fig:c40}). As another effect of the numerous
internal degrees of freedom of \alkane~undergoing 
conformational dynamics on multiple timescales, we have
identified a stretched exponential long-time decay of the friction
memory kernel. Indeed, the timescales of the conformational dynamics
of polymer chains have been linked to shear thinning \cite{Huber14},
with fast shearing leading to disentanglement of side chains and
decreasing fluid internal friction \cite{Datta21}. Since the kernel
decays too slow to be completely sampled by fast pulling velocities,
the lubricant moreover exhibits a generic velocity dependence as
discussed by Eq.~\eqref{eq:Gv}.
These findings are in line with microrheological experiments
that suggest a connection between the time scales of
external driving and the equilibrium relaxation time of a fluid
for the onset of non-Newtonian friction \cite{GomezSolano14}.

In summary, we have found that both long-time tails of correlation
functions and structural changes of the liquid may cause the
velocity-dependence of the friction in fluids. While previous
experimental and theoretical investigations proposed either of these
effects as the cause for non-Newtonian friction in selected fluids, we
have demonstrated that both effects are a general feature of fluids
independent of its internal complexity.

\begin{acknowledgements}
	
	We thank Kerstin Falk and Benjamin Lickert for numerous
	instructive and helpful discussions. This
	work has been supported by the Deutsche Forschungsgemeinschaft (DFG)
	via the Research Unit FOR 5099 ''Reducing complexity of
	nonequilibrium'' (project No. 431945604). The authors acknowledge
	support by the bwUniCluster computing initiative, the High Performance
	and Cloud Computing Group at the Zentrum f\"ur Datenverarbeitung of
	the University of T\"ubingen, and the Rechenzentrum of the University
	of Freiburg, the state of Baden-W\"urttemberg through bwHPC and the
	DFG through grants No.~INST 37/935-1 FUGG and No.~INST 39/963-1 FUGG.
	
\end{acknowledgements}

\bibliography{md,stock,new}

\end{document}


\author{Matthias Post}
\author{Steffen Wolf}
\email[email:~]{steffen.wolf@physik.uni-freiburg.de}
\author{Gerhard Stock}
\email[email:~]{stock@physik.uni-freiburg.de}
\affiliation{Biomolecular Dynamics, Institute of Physics, Albert Ludwigs
University, 79104 Freiburg, Germany.}
\title{Supporting Information: Molecular origin of driving-dependent friction in fluids} 
\date{\today}

\maketitle

\baselineskip5mm 
%
%

\section{Supplemental Methods}

In the main text, the simple force auto-correlation model
\begin{align}
\left< \delta f(t^\prime) \delta f(t^\prime{}^\prime) \right> = \frac{\gamma}{\beta \tau_C} \euler^{-\left|t^\prime -t^\prime{}^\prime\right|/\tau_C} 
\end{align}
was discussed, leading with Eq.~(6) and (22) to a velocity dependent averaged friction of
\begin{align}
\widehat{\itGamma}(v) =  \gamma \left( 1- \frac{v}{v_C} + \frac{v}{v_C} \, \euler^{-\frac{v_C}{v}}\right) \, .
\end{align}

If we model a system with two characteristic time-scales, we get
\begin{subequations}
	\begin{align}
	\left< \delta f(t^\prime) \delta f(t^\prime{}^\prime) \right> &= A_1 \, \euler^{- \frac{\left|t^\prime -t^\prime{}^\prime\right|}{\tau_1}} + A_2 \, \euler^{- \frac{\left|t^\prime -t^\prime{}^\prime\right|}{\tau_2}}  \\
	\widehat{\itGamma}(v) &=  \gamma \, \left( 1- \tfrac{v \widehat{\tau}}{x-x_0} + \tfrac{v \widehat{\tau}_1}{x-x_0} \, \euler^{-\frac{x-x_0}{v \tau_1}} + \tfrac{v \widehat{\tau}_2}{x-x_0} \, \euler^{-\frac{x-x_0}{v \tau_2}} \right) \label{eq:model_two_exp} \\
	\text{with} \quad \gamma &= \beta ( A_1 \tau_1 + A_2 \tau_2 ) \; , \quad \widehat{\tau} = \frac{A_1 \tau_1^2 + A_2 \tau_2^2 }{A_1 \tau_1 + A_2 \tau_2} \; , \nonumber \\
	\widehat{\tau}_{1} &= \frac{A_{1} \tau_{1}^2 }{A_1 \tau_1 + A_2 \tau_2} = \frac{\tau_2 - \widehat{\tau}}{\tau_2 - \tau_1} \, \tau_1 \; , \quad \widehat{\tau}_{2} = \frac{A_{2} \tau_{2}^2 }{A_1 \tau_1 + A_2 \tau_2} = \frac{\widehat{\tau} - \tau_1}{\tau_2 - \tau_1} \, \tau_2 \nonumber \; .
	\end{align}
\end{subequations}

Lastly, auto-correlation functions of more complex fluids are known to show the behavior of a stretched-exponential, which can be explained by an interplay of a continuous distribution of timescales, $\euler^{-(t/\tau)^\alpha} = \int_0^\infty \diff \tau^\prime c(\tau^\prime) \euler^{-t/\tau^\prime}$. This ansatz leads to
\begin{align}
\left< \delta f(t^\prime) \delta f(t^\prime{}^\prime) \right> &= A \, \euler^{- \left( \frac{\left|t^\prime -t^\prime{}^\prime\right|}{\tau}\right)^{\!\alpha} }\, , \label{eq:model_longtimetail}\\
\widehat{\itGamma}(v) &= \frac{\beta A \tau}{\alpha} \left[ \gpi\left(1/\alpha,\left( v_C/v \right)^\alpha\right) - \frac{v}{v_C} \, \gpi\left(2/\alpha,\left( v_C/v \right)^\alpha\right) \right] \, ,
\end{align}
where $\gpi(a,b) = \int_0^b \diff t \, t^{a-1} \euler^{-t}$ is the lower incomplete gamma function.

\clearpage
%

\section{Supplemental Results}

\begin{figure}[h!]
\begin{centering}	
\includegraphics[width=0.9\textwidth]{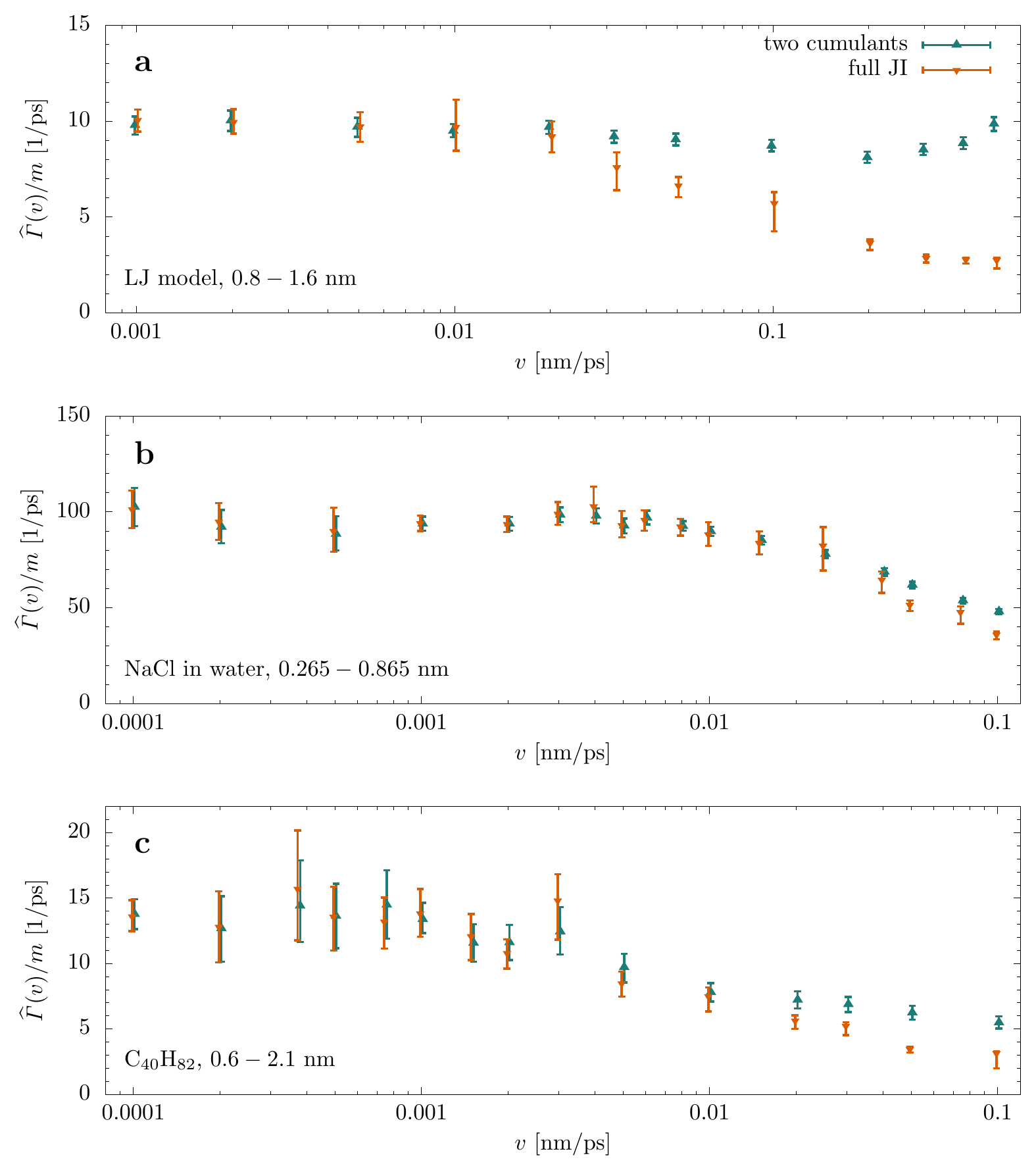}
\caption{Comparison of the position-averaged friction [Eq.~(22)] as a function of the pulling velocity $v$, using the cumulant approximation [Eq.~(6)] and the full exponential estimator [Eq.~(10)]. Shown are the results for (a) the Lennard Jones (LJ) model, (b) solvated NaCl, and (c) C$_{40}$H$_{82}$. While for low velocities the two estimates agree, they may differ for larger $v$. Here, the exponential estimate goes down faster in all cases, but most notably for the LJ model (a), despite being the simplest example of the three. The estimate of the dissipative work is higher for the cumulant approximation than using all moments, translating in higher friction than for the estimate with worse convergence behavior. The error bars indicate 95\% confidence intervals estimated by bootstrapping.}
\label{fig:SI_FIG1}
\end{centering}
\end{figure}

\clearpage

\begin{figure}[h!]
	\begin{centering}	
		\includegraphics[width=0.9\textwidth]{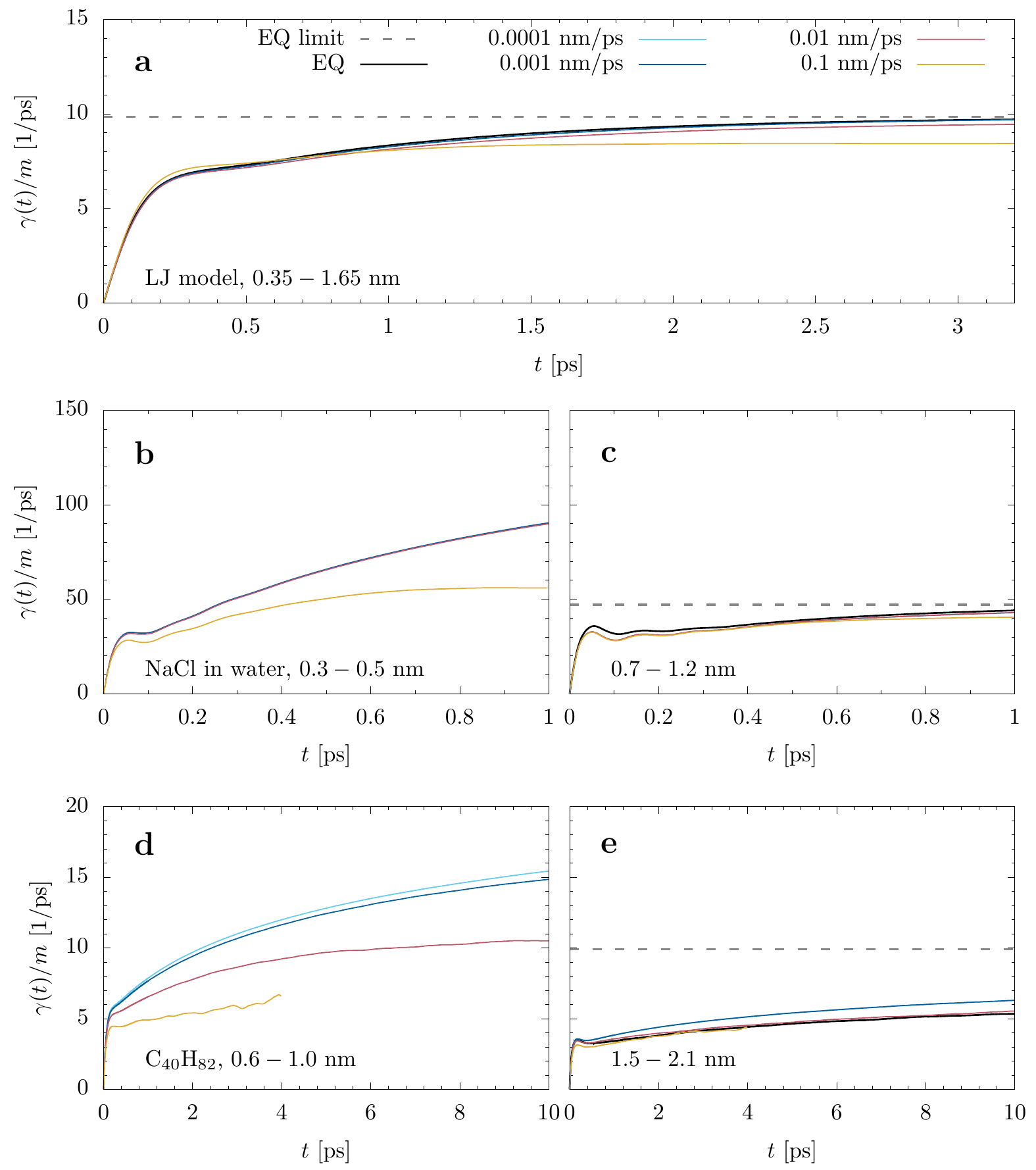}
		\caption{Convergence of friction estimate from integrating the memory kernel $\gamma(t)=\int_0^t \mathrm{d} \tau K(\tau)$, comparing results from unbiased and pulling simulations [Eqs.~(20) and (21)]. Estimates of the plateau values of the equilibrium kernel are shown in dashes. In more detail, to ensure a reasonable integration, we first needed to correct, in the case of Argon and NaCl, the artificial force spike at $t\!=\!0$ due to the 10 fs time-step by employing a parabolic fit to the subsequent time frames.  Moreover, since the iterative estimator Eq.~(21) appeared to fail at longer times, we performed a fit of the tail to correct for the noise and to extrapolate at longer times. That is, for the Lennard-Jones model we fit a mono-exponential function to the results obtained for 0.8-1.4~ps, in order to extend the kernel	from 1.4~ps onward. For NaCl we performed an exponential fit from 0.4 to 2.0 ps to connect at 0.5~ps, and for C$_{40}$H$_{82}$ we used a stretched-exponential fit, Eq.~(S4), from 2.2 to 20~ps (resulting in $\alpha \approx 0.2$) to continue from 10~ps. These fits yield a clear plateau value, although for C$_{40}$H$_{82}$ this procedure might still introduce a large systematic error due to a strong dependence of the parametrization of the long-time tail.		
		(a) For Argon, $\gamma$ converges after a few picoseconds and thus, even for fast pulling velocities, the generic convergence behavior is no issue here. 
		For NaCl, we compare $\gamma(t)$ at close (b) and far (c) ion distances. In the former case, the friction takes longer to converge to a higher value, but is still too fast to cause a generic effect. Instead, for fast $v$, $\gamma(t)$ converges faster and to a smaller value, explaining the decrease in Fig.~(3) in the main text. In the latter case, we can compare the estimate to the kernel of the unbiased simulation, which overall tend to agree and again converge to the same friction coefficient. For C$_{40}$H$_{82}$ (d,e), again, for close $x$, the plateau value of $\gamma$ is much higher than for large $x$, with slow, non-exponential convergence, thus potentially having a generic influence to the friction estimate of the main text. However, for fast $v$, there is again a faster decay to a smaller value. The generic convergence issue plays therefor again only a minor role.}
		\label{fig:SI_FIG2}
	\end{centering}
\end{figure}

\clearpage

\begin{figure}[h!]
	\begin{centering}	
		\includegraphics[width=0.9\textwidth]{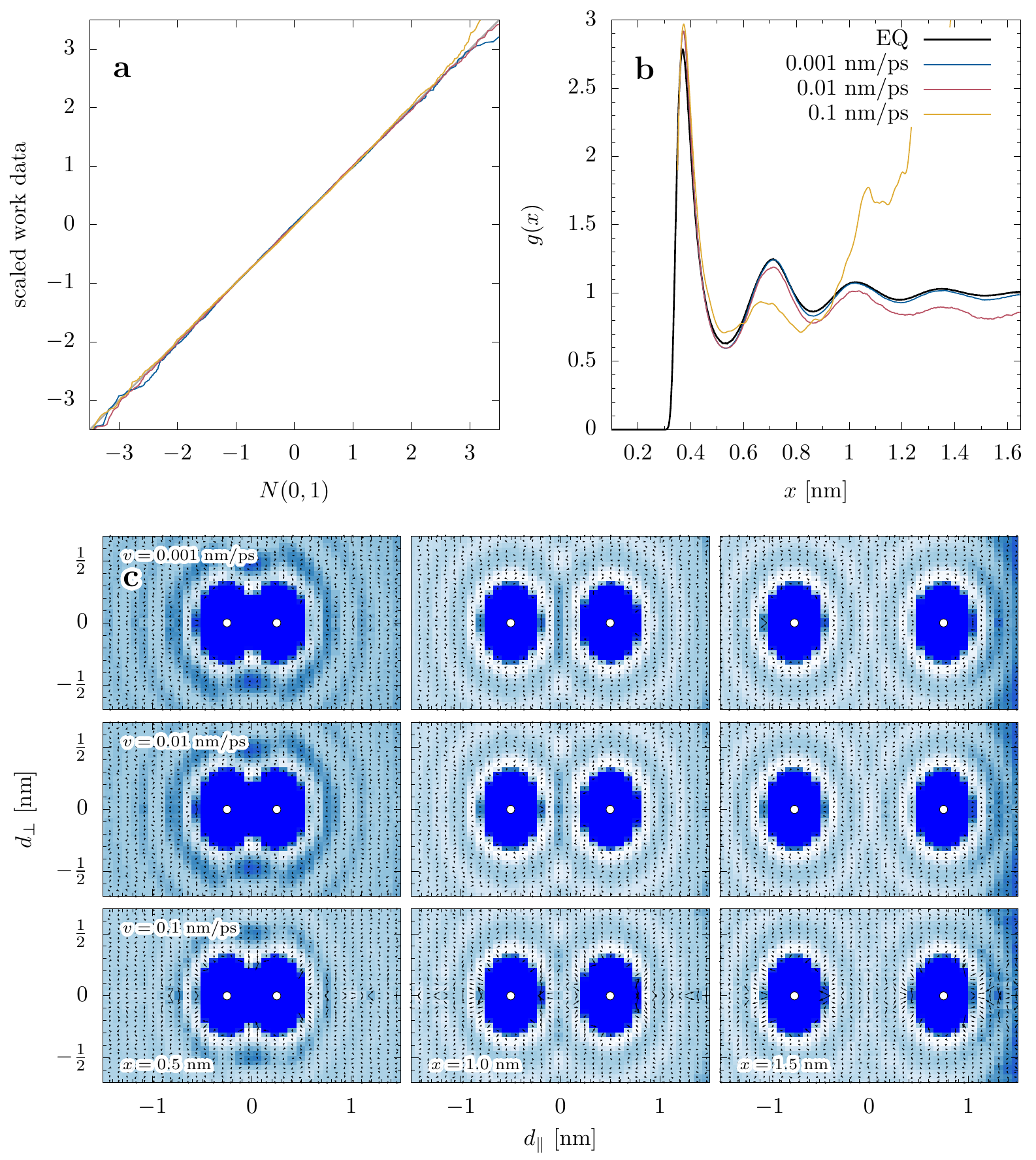}
		\caption{Supplemental information for the Lennard Jones liquid. (a) Normal probability plot of work distribution at $x=1$ nm for different pulling velocities, showing that the distribution follows a normal shape up to noise at the very tails. (b) Radial distribution function estimated in the unbiased case by gmx rdf and using Jarzynski's equality for the pulling simulations via $g(x)\propto\euler^{-\beta \Delta G(x)}/x^2$. (c) Distribution of surrounding bath Argon atoms with respect to the distance vector of the two probe particles (white=high density, blue=small density). The arrows indicate the average parallel and perpendicular velocities of the bath particles to exclude influences of possible eddies or flows. The picture is mirrored at the horizontal axis.}
		\label{fig:SI_FIG3}
	\end{centering}
\end{figure}

\clearpage

\begin{figure}[h!]
	\begin{centering}	
		\includegraphics[width=0.9\textwidth]{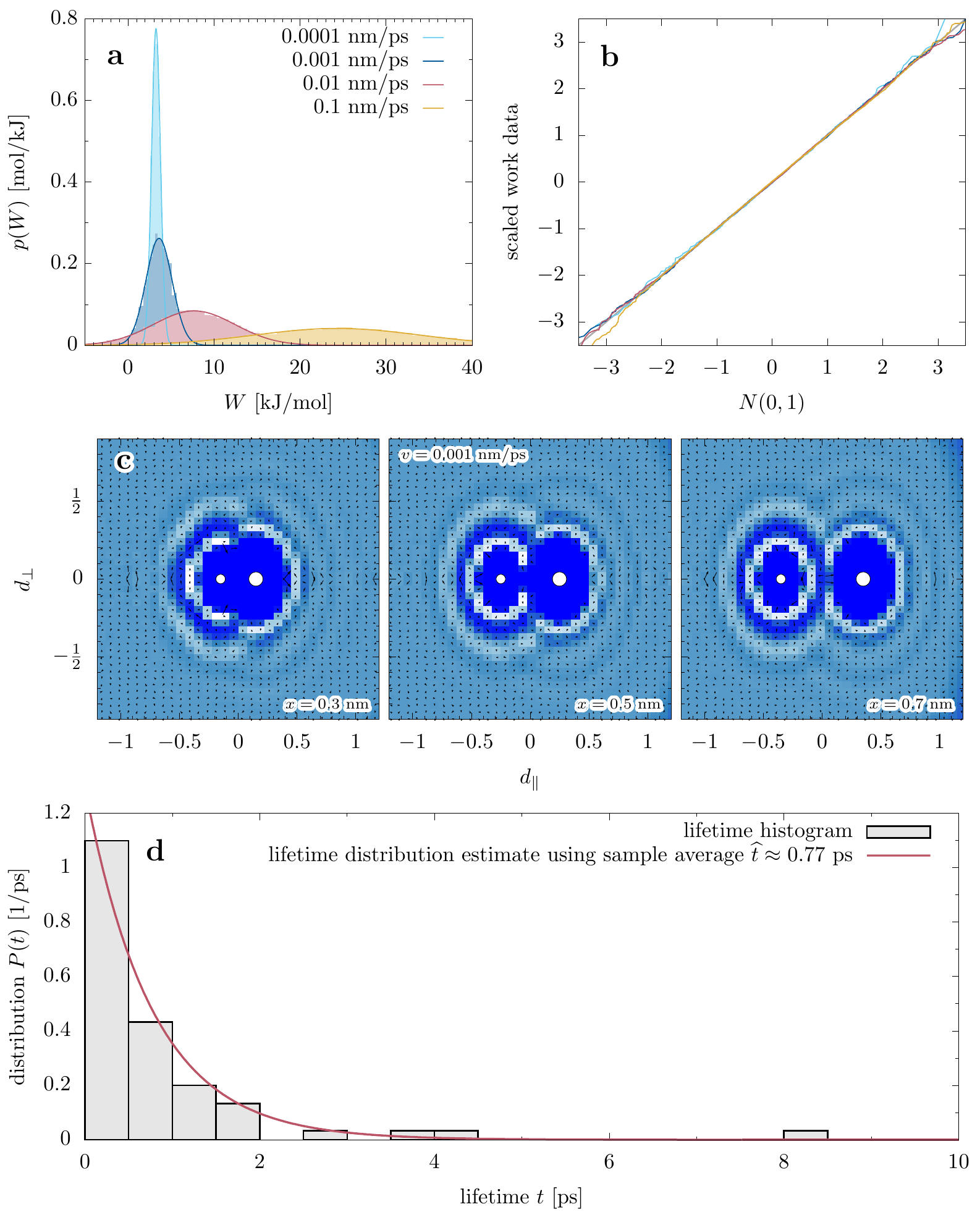}
		\caption{Supplemental information for NaCl in water. (a,b) Work distribution at 0.53~nm for different pulling velocities. While the distribution clearly gets broader, it still follows a normal shape up to noise at the very tails. (c) Distribution of surrounding water oxygen atoms with respect to the distance vector of the two ions (white = high density, blue = small density). The arrows indicate the average parallel and perpendicular velocities of the bath particles to exclude influences of possible eddies or flows. The picture is mirrored at the horizontal axis. (d) Lifetime distribution of waters inside the first shell.}
		\label{fig:SI_FIG4}
	\end{centering}
\end{figure}

\clearpage

\begin{figure}[h!]
	\begin{centering}	
		\includegraphics[width=0.9\textwidth]{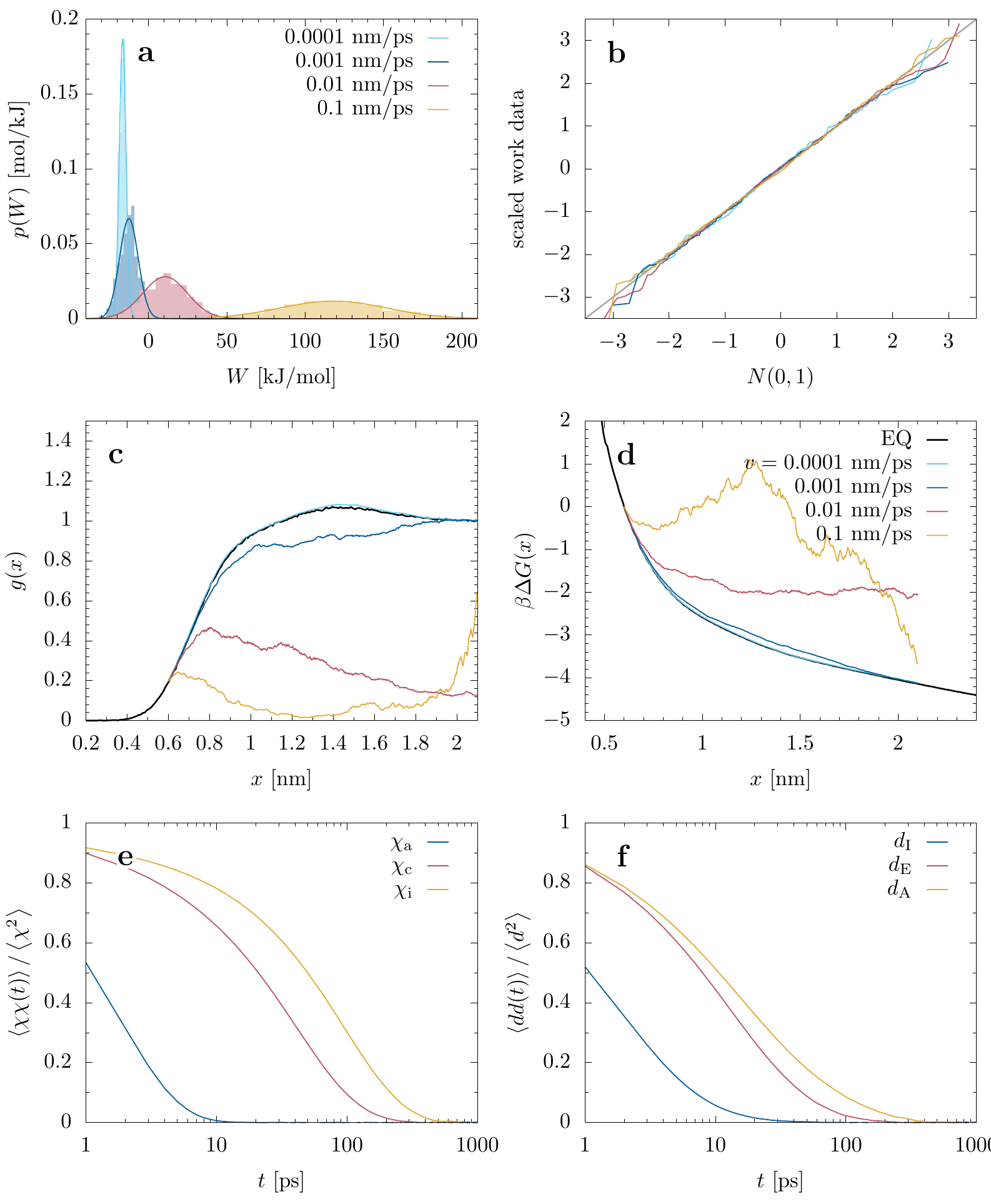}
		\caption{Supplemental information to C$_{40}$H$_{82}$ bulk liquid. (a,b) Work distribution at $x=1.4$~nm for different pulling velocities. While the distribution clearly gets broader, it still follows a normal shape up to noise at the very tails. (c,d) distribution of $x$ as radial distribution and free energy, respectively. Due to the complex shape of the alkanes, the estimate becomes poor already for moderate pulling velocities due to poor convergence of Jarzynski's equality. However, since we belief that the work variance itself is estimated properly, we may still trust the resulting friction (and the statistical error visualized by the error bars, e.g.~in Fig.~\ref{fig:SI_FIG5}). (e,f) Auto-correlation of internal coordinates of individual alkanes parametrizing their shape. While there are no clear collective motions of the arms, we can still estimate the time-scales of individual movements, e.g., that of the dihedral angles at the center and the tips (see Fig.~4a of the main text), as well as the arm distances, which happen from a few to hundreds of picoseconds.}
		\label{fig:SI_FIG5}
	\end{centering}
\end{figure}